\documentstyle[aps,prl,preprint,floats,epsfig]{revtex}

\def\LC{\Lambda_c^+} 
\def\LA{\Lambda^0} 
 
\def\EP{e^{+}}

\def\NE{\nu_e}

\def\To{\rightarrow}

\def\LCPSL{\LC \To \LA \EP \NE} 
 
\def\OMCSL{\Omega_c^0 \To \Omega^- \EP \nu_e} 
\def\CASCZSL{\Xi_c^{0} \To \Xi^{-} \EP \NE} 
\def\CASCPSL{\Xi_c^{+} \To \Xi^{0} \EP \NE}

\begin{document}

\preprint{\tighten\vbox{
	 	\hbox{\hfil CLNS 02/1782 }
		\hbox{\hfil CLEO 02~--~05 }}}

\title{Observation of the Decay ${\boldmath \Omega^{0}_c \rightarrow \Omega^{-} e^+ \nu_{e}}$ }

\author{CLEO Collaboration}

\date{June 5, 2002}

\maketitle

\tighten

\begin{abstract} 

Using the CLEO detector at the Cornell Electron Storage Ring 
we have observed the $\Omega_c^0$ ($css$ ground state) 
in the decay $\Omega_c^0 \rightarrow \Omega^- e^+ \nu_{e} $.  
We find a signal of $11.4 \pm 3.8 {\rm ~ (stat)}$ events. 
The probability that we have observed a 
background fluctuation is $7.6 \times 10^{-5}$.
We measure
$\ {B}(\Omega^{0}_{c} \rightarrow \Omega^{-} e^+ \nu_{e} ) \cdot 
\sigma(e^+ e^- \rightarrow \Omega^{0}_{c} X ) = (42.2 \pm 14.1 {\rm (stat)}   \pm 5.7 {\rm (syst)})$~fb
and $
 R =  \frac{ \Gamma (\Omega^{0}_c \rightarrow \Omega^{-} \pi^+) } {\Gamma ( \Omega^{0}_c 
\rightarrow 
\Omega^{-} e \nu_{e} )} = 0.41 \pm 0.19 {(\rm stat)}   \pm 0.04 {(\rm syst)}$.
This is the first statistically significant observation of an 
individual decay mode of the $\Omega^{0}_c$ in $e^+e^-$ annihilation,
and the first example of a baryon decaying via $\beta$~-~emission, where
no quarks from the first generation participate in the reaction.

\end{abstract}

\newpage

\tighten

\begin{center}
R.~Ammar,$^{1}$ D.~Besson,$^{1}$ X.~Zhao,$^{1}$
S.~Anderson,$^{2}$ V.~V.~Frolov,$^{2}$ Y.~Kubota,$^{2}$
S.~J.~Lee,$^{2}$ S.~Z.~Li,$^{2}$ R.~Poling,$^{2}$ A.~Smith,$^{2}$
% ...
\end{center}
 
\small
\begin{center}
$^{1}${University of Kansas, Lawrence, Kansas 66045}\\
$^{2}${University of Minnesota, Minneapolis, Minnesota 55455}\\
$^{3}${State University of New York at Albany, Albany, New York 12222}\\
$^{4}${Ohio State University, Columbus, Ohio 43210}\\
$^{5}${University of Oklahoma, Norman, Oklahoma 73019}\\
$^{6}${University of Pittsburgh, Pittsburgh, Pennsylvania 15260}\\
$^{7}${Purdue University, West Lafayette, Indiana 47907}\\
$^{8}${University of Rochester, Rochester, New York 14627}\\
$^{9}${Southern Methodist University, Dallas, Texas 75275}\\
$^{10}${Syracuse University, Syracuse, New York 13244}\\
$^{11}${University of Texas - Pan American, Edinburg, Texas 78539}\\
$^{12}${Vanderbilt University, Nashville, Tennessee 37235}\\
$^{13}${Wayne State University, Detroit, Michigan 48202}\\
$^{14}${California Institute of Technology, Pasadena, California 91125}\\
$^{15}${University of California, San Diego, La Jolla, California 92093}\\
$^{16}${University of California, Santa Barbara, California 93106}\\
$^{17}${Carnegie Mellon University, Pittsburgh, Pennsylvania 15213}\\
$^{18}${Cornell University, Ithaca, New York 14853}\\
$^{19}${University of Florida, Gainesville, Florida 32611}\\
$^{20}${Harvard University, Cambridge, Massachusetts 02138}\\
$^{21}${University of Illinois, Urbana-Champaign, Illinois 61801}\\
$^{22}${Carleton University, Ottawa, Ontario, Canada K1S 5B6 \\
and the Institute of Particle Physics, Canada M5S 1A7}
\end{center}

\newpage

The transition rate for charm quark semileptonic decays
is determined by the Cabibbo-Kobayashi-Maskawa matrix elements $|V_{cd}|$ and 
$|V_{cs}|$ and heavy quark form factors. Since both $|V_{cd}|$ and $|V_{cs}|$ are known
from three generation unitarity, measurements of charm semileptonic decays
allow an absolute measurement of the form factors~\cite{PDG2000}.

Within heavy quark effective theory (HQET)~\cite{HQET},
$\Lambda$-type baryons are more straightforward to treat than mesons as they consist of
a heavy quark and a spin and isospin zero light diquark. This 
simplicity allows for more reliable predictions for heavy quark to light quark 
transitions~\cite{LONG} than in the case for mesons. For example,
the measurement of the form factors
in $\LCPSL$ aids the future determination of 
$|V_{ub}|$ and $|V_{cb}|$ using $\Lambda^{0}_b$ decays since HQET relates the 
form factors in $\Lambda_c^+$ decay to those governing $\Lambda^{0}_b$ decays.

However,  it is important to test the theoretical treatment of charm 
baryon semileptonic decays. 
In this letter we report the first 
observation of $\Omega_c^0 \rightarrow \Omega^{-} e^+ \nu_{e} $. 
The $\Omega_c^0$  $(c\{ss\})$ is a 
$J^P = 1/2^+$ ground state baryon where $\{ss\}$ denotes the symmetric nature
of its wave function with respect to the interchange of the light-quark spins.
As $\OMCSL$ 
is a $J^P = 1/2^+ \rightarrow 3/2^+$ transition it is sensitive to additional form factors
not present in  $\LCPSL$, and so provides new information to test theory~\cite{Review}.

The data sample used in this analysis 
was collected with CLEO II~\cite{detector} and the upgraded CLEO II.V~\cite{det2}
detector operating 
at the Cornell Electron Storage Ring (CESR). The integrated luminosity 
consists of  
13.75 fb$^{-1}$ taken at and just below the $\Upsilon (4S)$ resonance
corresponding to approximately 18 million  $e^+e^-\rightarrow 
c\overline{c}$ events.

We search for the decay $\Omega^{0}_c \rightarrow \Omega^{-} e^+ \nu_{e}$ in 
$e^+e^-\rightarrow  c\overline{c}$ events
by detecting an $\Omega^- e^+$ (Right Sign) pair with 
invariant mass in the range 
$m_{\Omega^{-}} < m_{\Omega^{-} e^+} < m_{\Omega^{0}_c}$~\cite{charge}. 
The technique is very similar to that used in previous 
CLEO analyses of $\LCPSL$, $\CASCZSL$ and $\CASCPSL$~\cite{lambdaCStudy,cascadeCStudy,ffstudy}.

Positrons are identified using a likelihood function which incorporates
information from the calorimeter and $dE/dx$ systems. We require the positron to satisfy
$ | \cos \theta | < 0.71$  where $\theta$ is the 
angle between the positron momentum and the beam line. 
A positron is also required to 
originate from the primary vertex and have a momentum greater than 0.5~GeV/$c$.    
Muons are not used as  $\Omega^{0}_c \rightarrow \Omega^{-} l^+ \nu_{l}$ 
produces predominantly low momentum leptons and the CLEO muon identification system
is not efficient below 1~GeV/$c$.

The $\Omega^{-}$ is reconstructed in 
the decay $\Omega^{-} \rightarrow \Lambda^{0} K^{-}$,
$\Lambda^{0} \rightarrow p^+ \pi^{-}$.  
The analysis procedure for reconstructing these particles
closely follows that presented elsewhere~\cite{lambdaCStudy,cascadeCStudy,ffstudy,Basit,early_work_1}. 
Kaon and proton candidates must have
specific ionization and time-of-flight measurements consistent with the expected
values. Particle identification is not used for pions.
The hyperons are required to have vertices well separated from the beam spot, with the
flight distance of the secondary $\Lambda^{0}$ greater than that of the $\Omega^{-}$.
The $\Omega^{-}$ is required to originate from the primary vertex of
the event. To reduce background in $\Omega^{-}$ reconstruction,
kaons and $\Lambda^{0}$'s
consistent with originating from the primary vertex are excluded.
In order to improve mass resolution in the $\Omega^{-}$ reconstruction 
the $\Lambda^{0}$ mass constraint is employed.

The Monte Carlo (MC) simulated signal events used in this analysis were generated
for the two detector configurations using a GEANT-based~\cite{GEANT}
simulation and were processed similarly to the data. We take 
$m_{\Omega^{0}_c}=2704.0$ MeV/$c^2$~\cite{PDG2000}.  
For $\Omega^{0}_c \rightarrow \Omega^{-} e^+ \nu_{e}$
we assume no net polarization of the $\Omega^{-}$~\cite{Om_model}.
The fragmentation function for the $\Omega^{0}_c$ is unknown, 
therefore the  measured fragmentation function of 
the $\Xi_c$~\cite{fragFunction} is used.

\begin{figure}[h]
   \begin{center}
      \epsfig{figure=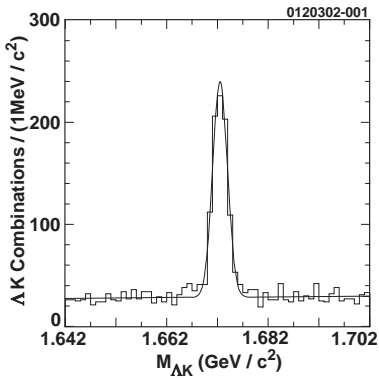,width=2.0in,height=2.0in}
   \end{center}
   \caption{Invariant mass of $\Lambda^{0} K^-$ combinations. }
   \label{fig:set1A}
\end{figure}

Figure~\ref{fig:set1A}
shows the invariant mass distribution of $\Lambda^{0} K^-$ pairs
with all selection criteria imposed.  The signal is fit by a Gaussian  
and the background is parameterized 
by a second order polynomial function.  
The signal yield from the fit is  $763 \pm 32 $. 
The mean and width of the Gaussian   
are $1672.50 \pm 0.07$~MeV/c$^2$ and  
is $1.44 \pm 0.06$~MeV/$c^2$ respectively. The width
is consistent with that expected from MC simulation.

\begin{figure}[h] 
   \begin{center} 
      \epsfig{figure=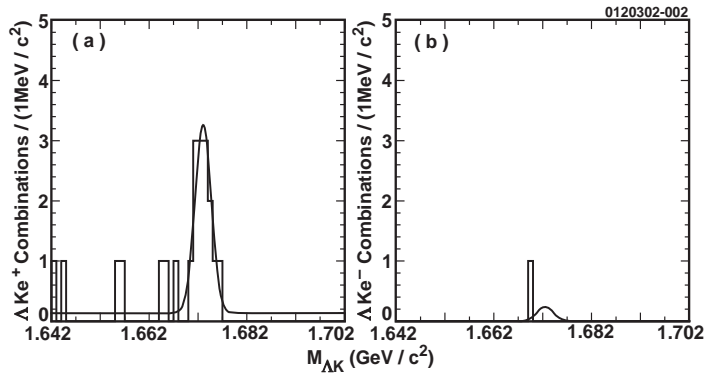,width=4.0in,height=2.0in}
   \end{center}
   \caption{The invariant mass of $\Lambda^{0} K^-$ pairs for events 
with a positron (right plot) and an electron (left plot)
satisfying the selection criteria described in the text.}
   \label{fig:set2BC} 
\end{figure}

The $\Omega^{-}$ candidates are combined with positrons. The invariant 
mass of the $\Omega^{-} e^+$ pair 
is required to satisfy $ m_{\Omega^{-}} < m_{\Omega^{-} e^+} < m_{\Omega^{0}_c}$. 
We require $|\vec{p}_{\Omega^{-}} +  \vec{p}_{e^+}|> 1.4 {\rm~GeV}/c$ to 
reduce background from $B \overline{B}$ events. 
Figure~\ref{fig:set2BC} 
shows the invariant mass distributions of $\Lambda^{0} K^-$ pairs in events
which contain 
a right sign (RS) or wrong sign (WS) positron. There is a pronounced excess of 
RS events compared to WS events at the $\Omega^-$ mass  as 
would be expected if we are observing the decay 
$ \Omega^{0}_{c} \rightarrow \Omega^{-} e^+  \nu_{e} $. 
The $\Lambda^{0} K^-$ invariant mass distributions 
are fit with a function consisting of a Gaussian with width determined by a MC simulation
to represent the signal and a first order polynomial function to represent the background.
We define the signal box as a $\pm 3.0\sigma $ signal region around the $\Omega^{-}$ mass.
The fit returns $13.0 \pm 3.8$  $(1.3 \pm 0.6)$  events in the Gaussian component
(background component) within the signal box.

We now consider backgrounds to the signal.
There are four types of background that produce events
that populate the signal box.
These are:  (1) fake $e^+$ -- real $\Omega^{-}$ combinations, 
(2) random  $\Omega^{-} e^+$ pairs from (a) the continuum 
(generic $e^+e^- \rightarrow q \overline{q} $ events) where  
the $\Omega^{-}$ is not a decay product of a charm baryon semileptonic decay
and (b) $B$ decays at the $\Upsilon (4S)$, (3) feeddown from decays of the type 
$\Omega^{0}_c \rightarrow \Omega^{-} X e^+ \nu_{e}$, 
where $X$ is an unobserved decay product,  (4) feedthrough from $\Xi_c$ and $\Lambda_c$ semileptonic 
decays.  In addition for each of (1), (2), (3), and (4) there is combinatorial 
background to the $\Omega^-$ (usually a $\Lambda^0$ 
with random kaon). This fake $\Omega^-$ background is uniformly distrubuted in the
$\Lambda^0 K^-$ invariant mass and populates both the signal box and the region
outside the signal box. Therefore the population of events outside the signal box can
be used to check etimates of the number of background events in the signal box.
A further check is provided by wrong sign events which are produced by (1) and (2).

The evaluation of the backgrounds is as follows.
The fake positron contribution to both RS and WS events 
depends on the particle populations in $c \overline{c}$ jets 
containing an $\Omega^{-},$ and the species and momentum-dependent fake 
rate\cite{refwhich}. 
In this analysis, strangeness (baryon number) conservation leads to enhanced
kaon (antiproton) production
in $e^+ e^- \rightarrow \Omega^{-} X$ events.   
Both antiprotons and kaons  have larger positron fake rates than 
pions. Because of baryon conservation, fake leptons from baryons
are much more numerous in WS than in RS combinations.
To account for the different positron fake rates 
of each particle species, all tracks in events containing an $\Omega^{-}$
that are not positively identified as positrons, 
are weighted by the momentum-dependent positron fake rates for each particle species, 
and the particle populations in $c \overline{c}$ jets containing an $\Omega^{-}$,
determined from data. We estimate there are $1.4 \pm 0.4$~($0.2 \pm 0.2$) 
fake $e^+$~--~real $\Omega^{-}$  RS events due to kaons and protons (pions) 
faking positrons in the signal box. Thus, the total contribution from this source
is $1.6 \pm 0.5$ RS events.

The $\Omega^{-}$ production mechanisms  in continuum events  
and $\Upsilon (4S) \rightarrow B \bar{B} $ decays are
not well known, therefore  
a MC estimation of random combinations of 
real $\Omega^{-} e^+$ pairs from these processes will be unreliable. However, 
previous CLEO analyses found that 
the background from random  $\Lambda^{0} e^+$ and $\Xi e^+$ pairs 
in the modes $\LCPSL$, $\CASCZSL$, and $\CASCPSL$ is 
small and is likely 
to populate both RS and WS equally~\cite{lambdaCStudy,cascadeCStudy,ffstudy}.
In this analysis the absence of WS events at 
the $\Omega^{-}$  mass demonstrates that the random
pairing of an $\Omega^{-}$ and a electron is negligible. We assume 
that this is also true for RS combinations.

Background due to decays of the type  $\Omega^{0}_c \To \Omega^{-}  e^+ \nu_{e} X$, 
for example,   $\Omega^{0}_c \To \Omega^{*-} e^+ \nu_{e} $, 
$\Omega^{*-} \To \Omega^- X$ produces a peak in the 
$\Lambda^{0} K^-$  mass distribution. 
The lightest and best understood resonance in the $\Omega^-$ family is 
the $\Omega (2250)^-$~\cite{PDG2000} which does not decay to 
an $\Omega^{-}$. 
The $\Omega(2470)^-$ decays to $\Omega^{-} \pi^+ \pi^-$,
however, because the mass of this resonance is close to the $\Omega^{0}_c$ mass, 
the phase space suppression will be severe, and the positron spectrum entirely 
below 0.5 GeV/$c$. We note that due to isospin conservation
the decay  $\Omega^{*-} \rightarrow \Omega^- \pi^0$ is forbidden. 
If a, yet to be discovered, $\Omega^{*-}$ with a mass in 
the range $m_{\Omega^-} + 2 m_{\pi} < m_{\Omega^{*-}} < 2.250$~GeV/$c^2$
exists it could, in principle, constitute a background to
this analysis through the decay $\Omega^{*-} \rightarrow \Omega^- (\pi \pi)^0$.
However, it is likely that the dominant decay would be  
$\Omega^{*-} \rightarrow \Xi K$ which has a larger
phase space available. Given that no light $\Omega^{*-}$ has been identified
we do not consider  this possibility further. We conclude that this analysis
is insensitive to $\Omega^{0}_c \To \Omega^{-}  e^+ \nu_{e} X$.

\begin{figure}[h]
   \begin{center}
      \epsfig{figure=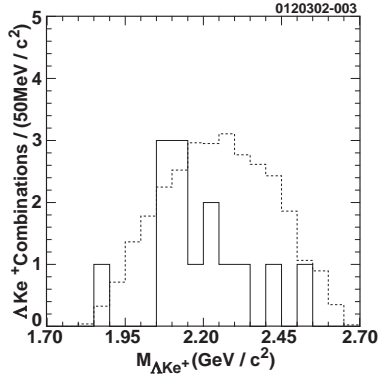,width=2.0in,height=2.0in}
   \end{center}
   \caption{$\Omega^- e^+$ invariant mass distribution for events in the signal 
 box (solid line) and a Monte Carlo simulation (dashed line).}
   \label{fig:candidatesSpectra}
\end{figure}

The modes $\Xi_{c}^+ \rightarrow \Omega^- K^+ e^+ \nu_{e}$  and 
$\Xi_{c}^0 \rightarrow \Omega^- K^0 e^+ \nu_{e}$ 
also produce a peak in the $\Lambda^{0} K^-$  mass distribution.
However, these decays are expected to be suppressed for the following reasons.
Semileptonic decays favor little hadronic fragmentation. A study of
$\Lambda_c^+ \rightarrow \Lambda^{0} e^+ \nu_{e}$,~\cite{lambdaCStudy}
found that 
$B(\Lambda_c^+ \rightarrow \Lambda^{0} e^+ \nu_{e}) / 
 B(\Lambda_c^+ \rightarrow \Lambda^{0} e^+ \nu_{e} X) > 0.85$ 
{\rm at 90 \% confidence level}. The same pattern is seen in charm mesons~\cite{DtoK}.
In $B$ meson semileptonic decays where the energy release is larger
there is only modest non-resonant production~\cite{PDG2000}. 
Also $\Xi_{c}\rightarrow \Omega^- K e^+ \nu_{e}$  
proceeds via the creation of an $s \bar{s}$ pair from the vacuum, which is suppressed
relative to light quark anti-quark pair creation from the vacuum.
There is no experimental evidence for $s \bar{s}$ pair creation
in semileptonic decays of $b$ and $c$ quarks. 
In addition, as $\Xi_c \rightarrow \Omega^- K e^+ \nu_{e}$
produces softer leptons than $\Omega^{0}_c \rightarrow \Omega^- e^+ \nu_{e}$,   
the reconstruction efficiency is an order of magnitude 
lower, and  $M_{\Omega^- e^+} < 1.98$~GeV/$c^2$ is satisfied. 
Figure~\ref{fig:candidatesSpectra}  shows the 
$\Omega^- e^+$ invariant mass distribution for events
in the signal box and compares it to the distribution expected for 
$\Omega_{c}^0 \rightarrow \Omega^- e^+ \nu_{e}$. The data is consistent with the simulation.
There is one event with $M_{\Omega^- e^+} < 1.98$~GeV/$c^2$ consistent 
with $\Xi_c \rightarrow \Omega^- K e^+ \nu_{e}$.
As this event could be either signal or background, it contributes 
a $0^{+1}_{-0}$ event uncertainty to the number of signal events.

Feedthrough from other charm baryon semileptonic 
decays, $\LCPSL$, 
$\CASCZSL$, and $\CASCPSL$  (coherent background), is a source of $\Lambda^{0}$ 
$e^+$ pairs, which, when combined with a random track 
in the event 
satisfying the kaon hypothesis, can mimic the signal. 
Coherent background is evaluated by generating
MC events according to a HQET consistent model~\cite{KK}
which was shown in~\cite{ffstudy} to describe the decay $\LCPSL$. 
The MC events are generated using the measured fragmentation functions of the 
$\Lambda_c^+$, $\Xi_c^+$, and $\Xi_c^0$.
Since the $e^+ e^-$  cross section for each process  
has been measured ~\cite{lambdaCStudy,cascadeCStudy,ffstudy},
a reliable prediction of the coherent background can be made.
We estimate that the coherent background 
contributes $3.5 \pm 1.9$ RS events
distributed uniformly in the 
range $1.642< m_{\Lambda^{0} K^-}<1.072~ {\rm GeV/c^2}.$ 
Note that the coherent background, and any other source of fake
$\Omega^-$ -- real or fake  $e^+$ 
is automatically taken 
into account by fitting
the $\Lambda^{0} K^-$ mass to determine the yield.

We now compare our estimate of the RS and WS backgrounds to the data.
We estimate that fake positron 
background contributes $0.3 \pm 0.3$ ($0.4 \pm 0.4$) WS events 
in the signal box (outside the signal box). 
The sum is in good agreement with the one WS event observed.
We estimate coherent background (fake positron - fake $\Omega$) contribute
 $2.9 \pm 1.6$
($1.6 \pm 0.5$) RS events outside the signal box in reasonable agreement
with the the 7 RS events observed outside the signal box. The slight excess observed
in data may be attributed to
additional sources of fake $\Omega$ -- real $e^+$ pairs which have not been accounted for.
For example, the assumption that random pairs populate RS and WS equally may not be exact.
However, we remind the reader that, due to the fit,  
the excess is accounted for in the determination of the yield.
Finally, we estimate that the coherent background  
contributes $0.6 \pm 0.3$ events in the 
signal box in good agreement with the $1.3 \pm 0.6$ background events in the signal box 
returned from the fit.

To estimate the number of signal events in the signal box we subtract
the fake positron background from the Gaussian component of the fit
to obtain $11.4 \pm 3.8 $ events
consistent with $\Omega^{0}_c \rightarrow \Omega^{-} e^+ \nu_{e}$. 
The probability
for the background in the signal box ({\it i.e.} the sum of the coherent background
and the fake positron background) to fluctuate to 14 or more events
is  $2.3 \times 10^{-6}$.  Correcting the number of signal events by the
signal efficiency and integrated luminosity of the data sample our 
measured ${B} ( \Omega^{0}_c \rightarrow \Omega^{-} e^+ \nu_{e} ) 
\cdot \sigma (e^+ e^- \rightarrow \Omega^{0}_c X)$ is $(42.2 \pm 14.1  \pm 5.7)~~{\rm fb}$.

We have considered the following sources of systematic uncertainty and give our estimate 
of their magnitude in parentheses.
Background from the process $\Xi_c \rightarrow \Omega^{-} K e^+ \nu_{e}$ is
estimated from the $\Omega^{-} e^+$ invariant mass distribution  of 
Figure~\ref{fig:candidatesSpectra} (7.1\%). 
The uncertainty in the fake positron background is determined from our
knowledge of the species and momentum dependent fake rates and particle 
populations in $c \overline{c}$ jets containing an $\Omega^{-}$~(6.7\%).
The uncertainty associated with imperfect knowledge of the $\Omega^{0}_c$ 
fragmentation function is estimated by varying this function~(6.0\%).
The uncertainty associated with the baryon finding efficiency is determined 
by data and MC studies for the  $\Omega^{-}$ and $\Lambda^{0}$ to be
(5.0\%) and (4.0\%) respectively. This uncertainty includes the uncertainty 
associated with track finding efficiency for $p, \pi$ and $K$.
The uncertainty in finding the positron track is determined by our
knowledge of the the track finding efficiency of the CLEO II/II.V 
detectors~(1.0\%).
The uncertainty associated with the positron 
identification efficiency is determined by Bhabha embedding studies~(2.0\%).
The uncertainty associated with MC modeling of long-lived hyperons is 
estimated to be~2.0\%.
The uncertainties associated with MC modeling of slow pions from $\Lambda^{0}$ decays
is obtained by varying this efficiency according to our understanding
of the CLEO detector~(1.2\%). 
There is a 1.0\% systematic uncertainty
in the total integrated luminosity. 
The uncertainty in $B(\Omega^{-} \rightarrow \Lambda^{0} K^-)$ and 
$B( \Lambda^{0} \rightarrow p^+ \pi^-)$  contribute a 1.3\% uncertainty to 
our measurement. Finite MC statistics contribute a 0.1\% uncertainty
to the signal efficiency.  
The uncertainty in the efficiency associated with the choice 
of model for  the decay 
is estimated by comparing the efficiency with a matrix 
element producing $\Omega^{-}$'s with no net polarization (which is the
efficiency used for the result) and full polarization.
The difference in reconstruction efficiency for the two models
is negligible and no uncertainty is assigned from this source. Adding all sources
of systematic
uncertainty in quadrature, the total systematic uncertainty is found to be~13.6\%.

We compute the combined statistical and systematic significance 
of our observation by the following procedure. Most of the quantities for which
a systematic uncertainty has been assigned do not contribute to the uncertainty 
in the magnitude of the background, the exception is the uncertainties associated with
$\Xi_c \rightarrow \Omega^{-} K e^+ \nu_{e}$ and fake positron background. 
We assume the event satisfying $M_{\Omega^- e^+} < 1.98$~GeV/$c^2$ is
background from  $\Xi_c \rightarrow \Omega^- K e^+ \nu_{e}$ and increase the background
by one event. We also increase the background by our uncertainty in the 
fake positron background.
The probability for the  background to fluctuate to 14 or more events in the signal
is $  7.6 \times 10^{-5}$.

At present there is no reliable normalization of the $\Omega^{0}_c$
branching ratios. As the rates for semileptonic decays are, in principle, 
simpler to calculate than hadronic decays, the ratio of our ${\cal B} \cdot \sigma$
to that for a hadronic mode will be useful for normalizing the hadronic 
scale once reliable theoretical predictions exist for the semileptonic modes. 
Therefore we measure $R = {\Gamma} (\Omega^{0}_{c} \rightarrow \Omega^{-} \pi^+) / 
{\Gamma} (\Omega^{0}_{c} \rightarrow \Omega^{-} e^+ \nu_{e} )$.
We search for $ \Omega^{0}_{c} \rightarrow \Omega^{-} \pi^+ $ using a set of selection
criteria very similar to those in the $\Omega^{0}_{c} \rightarrow \Omega^{-} e^+ \nu_{e} $
analysis~\cite{detail}. We find $14.1 \pm 4.3$ events consistent with the decay 
$ \Omega^{0}_{c} \rightarrow \Omega^{-} \pi^+ $~\cite{comparison}.
After correcting the yields by the efficiencies we compute
$
R = {\Gamma(\Omega^{0}_{c} \rightarrow \Omega^{-} \pi^+) } / {
\Gamma (\Omega^{0}_{c} \rightarrow \Omega^{-} e^+ \nu_{e} )} = 0.41 \pm 0.19{\rm (stat)} \pm 
0.04{\rm (syst)}.$
Most of the systematic uncertainties cancel in
forming the ratio. The largest remaining sources of systematic uncertainty 
are associated with the estimates of background for the semileptonic
mode. The corresponding ratio in $\Lambda^{+}_c$ ($\Xi_{c}^0$) decays is 
$0.44 \pm 0.09 {\rm (stat)} $ ($0.3 \pm 0.1  {\rm (stat + syst)}$).

In summary, we have reconstructed 14 $\Omega^- e^+$ pairs of which $11.4 \pm 3.8$ are 
consistent with the decay
$\Omega^{0}_c \rightarrow \Omega^- e^+ \nu_{e} $. 
The probability that we have observed a  background 
fluctuation is  $7.6 \times 10^{-5}$. 
Our measured ${\cal B} \cdot \sigma$  is $(42.2 \pm 14.1  \pm 5.7)~~{\rm fb}. $
This is the first statistically significant observation of an 
individual decay mode of the $\Omega^{0}_c$ in $e^+e^-$ annihilation
and the first example of a baryon decaying via $\beta$~-~emission, where
no quarks from the first generation participate in the reaction.
We have also measured $R = {\Gamma} (\Omega^{0}_{c} \rightarrow \Omega^{-} \pi^+) / 
{\Gamma} (\Omega^{0}_{c} \rightarrow \Omega^{-} e^+ \nu_{e} ) = 0.41 \pm 0.19 \pm 0.04$.

\suppressfloats

We gratefully acknowledge the effort of the CESR staff in providing us with
excellent luminosity and running conditions.
This work was supported by 
the National Science Foundation,
the U.S. Department of Energy,
the Research Corporation,
and the Texas Advanced Research Program.

\suppressfloats


\begin{thebibliography}{99}
\bibitem{charge}   Throughout this paper charge conjugate modes are implicitly included.
\bibitem{Om_model} 
HQET predicts that at $q^2 = 0$ the $\Omega^{-}$ emerges with zero polarization.
This prediction is independent of the values of the six independent form factors
in this decay. The $\Omega^{-}$ polarization varies with $q^2$, however
setting it equal to zero is an adequate approximation for this analysis.
\bibitem{refwhich} Fake rates are measured from tagged data samples of $\pi^+/\pi^-$,
$K^+/K^-$ and $p^+/\overline{p}^-$ tracks.
\bibitem{comparison}  
The $\Omega^{0}_{c} \rightarrow  \Omega^{-} \pi^+$
yield in this analysis is consistent with that reported in~\cite{Basit}.
\end{thebibliography}
\end{document}